\documentclass[10pt]{article}
\usepackage{epsfig,amsmath,amssymb,graphics}

\newcommand{\half}{\frac{1}{2}}

\newcommand{\hc}{\text{ h.c.}}

\newcommand{\LL}{\mathcal{L}}
\newcommand{\Tr}{\text{ Tr }}

\newcommand{\diag}{\text{ diag }}
\newcommand{\eff}{{\text{eff}}}
\newcommand{\identity}{{\rlap{1} \hskip 1.6pt \hbox{1}}}

\begin{document}

\begin{titlepage}
\begin{flushright}
HUTP-02/A015\\
hep-ph/0206023\\
\end{flushright}
\vskip 2cm
\begin{center}
{\large\bf Mooses, Topology, and Higgs}
\vskip 1cm
{\normalsize
\mbox{
\hspace{-0.3in}
Thomas Gregoire and Jay G. Wacker}\\
\vskip 0.5cm

Department of Physics, University of California\\
Berkeley, CA~~94720, USA\\
and \\
Theory Group, Lawrence Berkeley National Laboratory\\
Berkeley, CA~~94720, USA
\vskip .3cm

Jefferson Physical Laboratory\\
Harvard University\\
Cambridge, MA 02138\\
\vskip .3cm

}
\end{center}

\vskip .5cm

\begin{abstract}
New theories of electroweak symmetry breaking  have recently been
constructed  that stabilize the weak scale and do not rely upon 
supersymmetry.  In these theories the Higgs boson is a weakly coupled
pseudo-Goldstone boson.
In this note we study the class of theories that can be described
by theory spaces and show that the fundamental group of theory
space describes all the relevant classical physics in the
low energy theory. 
The relationship between the low energy physics and the topological properties
of theory space allow a systematic method for constructing 
theory spaces that give any desired low energy particle content and potential. 
This provides us with tools for analyzing and constructing new theories of 
electroweak symmetry breaking. 
\end{abstract}

\end{titlepage}

\section{Introduction}

The description of electroweak symmetry breaking in the Standard
Model, in terms of a fundamental scalar Higgs field, suffers from a
stability crisis.  
The quadratically divergent radiative corrections
to the Higgs mass suggest that the description of TeV scale
physics in the Standard Model is incomplete.
New physics at the TeV scale must emerge to stabilize the weak scale.  
Recently, a qualitatively new category of realistic theories of electroweak
symmetry breaking has been introduced \cite{Arkani-Hamed:2001nc}.
These models, based on deconstruction
\cite{Arkani-Hamed:2001ca,Hill:2000mu,Cheng:2001vd} and the physics of ``theory
space''\cite{Arkani-Hamed:2001vr,Arkani-Hamed:2001ed,Cheng:2001nh,Arkani-Hamed:2001ie} offer a new mechanism for
softening the quadratic divergences in the Higgs mass.  Electroweak
symmetry breaking is accomplished with naturally light Higgs bosons
that descend from non-linear sigma model fields whose mass is
protected by ``chiral'' symmetries of the sigma model. The first
attempts at models of this kind were the ``composite Higgs'' theories
\cite{Kaplan:1983fs,Kaplan:1983sm,Georgi:1984af} that required 
fine tuning to keep the Higgs light.  More
recently, models similar in spirit to the theory space models and using the
same group theory structure as the composite Higgs model
have been developed \cite{Coset}. 
In all of these theories, the physics is perturbative at
energies parametrically above the TeV scale, ultimately requiring an 
ultraviolet
completion near \mbox{$\sim 10$ TeV} where the non-linear sigma model
fields become strongly coupled. However, the physics of electroweak
symmetry breaking and the new physics at the TeV scale are weakly
coupled and do not depend on the ultraviolet completion. These models are fully realistic, incorporating fermion
masses without producing dangerous flavour-changing neutral currents
in the low energy theory.

The general structure of these models is characterized by a ``theory
space'', consisting of sites, lines and faces. Each site represents a gauge
group, each line represents a non-linear sigma model link field
transforming under the gauge groups at the ends of the line, and each 
face corresponds to ``plaquette'' operators involving a trace of
products of the link fields bounding the face. The little Higgs
descend from the link fields, while their quartic coupling arise
from the plaquette interactions. Based on deconstructing extra dimensional intuitions, the
models used in 
\cite{Arkani-Hamed:2001nc,Arkani-Hamed:2002pa} were $N \times N$
deconstructed torus. The basic
ingredients that make this class of models successful theory of electroweak
symmetry breaking are the absence of one loop quadratic divergences in the
Higgs mass, guaranteed by the approximate chiral symmetries, and the
presence of large quartic self interaction for the Higgs. 

In this paper, we seek a way of extracting the low energy physics of general
theory spaces in order to decide which spaces can be used for electroweak
symmetry breaking. We also develop a method for building theory spaces with
a given low energy particle content and potential. The spectrum of electroweak symmetry breaking
theories based on theory space is characterized by two or more Higgs doublets at
roughly  \mbox{100 GeV} and at least one TeV scale particle for each quadratic
divergence of the low energy theory. In contrast with supersymmetric
theories, quadratic divergences are canceled by `partners' of the same spin.

In Sec. \ref{Sec: Homotopy}, 
we review the structure of theory space and present a
systematic procedure to calculate the moduli space of general theory space,
allowing us to obtain the low energy potential of the theory. We illustrate
this procedure with several examples. 
We then reverse the logic and show how to build theory spaces that possess arbitrary low energy physics.

In Sec. \ref{Sec: Radiative Corrections}, we analyze the structure 
of radiative corrections
in little Higgs model, and present two simple rules that ensure that a theory
space is free of quadratic divergences at one loop. 
In Sec. \ref{Sec: Fermions} we discuss how to
include Yukawa couplings so that they do not reintroduce one loop
quadratic divergences.   We also show that it is possible for fermions
to generate the plaquette potential.

Finally, in Sec. \ref{Sec: Lifting} we discuss how to lift unnecessary
states out of the low energy theory and into the \mbox{1 TeV} range.
When the gauge symmetry is reduced at one site new plaquette potentials
are allowed that can differentiate between the adjoint states and
Higgs ( these are the $\mathbf{T_8}$ plaquettes of
\cite{Arkani-Hamed:2001nc,Arkani-Hamed:2002pa}). This allows us to build
models free of light triplet and singlet scalars that were present in other
little Higgs models constructed from theory space 
\cite{Arkani-Hamed:2001nc,Arkani-Hamed:2002pa,2sites}. In particular we
present an extension of the two sites model of \cite{2sites} where the $\sim 100$ GeV
triplet and singlet scalars of \cite{2sites} are pushed to the TeV scale.

\section{Topology and Theory Space}
\label{Sec: Homotopy}

There are general statements we 
can make about  the existence of little Higgs and their
potentials from the structure of theory space alone.  
Understanding the general structure of theory
space and its relation to the low energy dynamics will
allow us to classify the little Higgs theories and determine
if they are viable models of electroweak symmetry breaking.

The physics of little Higgs models is specified by the gauge structure, 
the link variables and the scalar potential, these define theory space 
by points, lines, and faces, respectively.  
Gauge groups are labeled by points: $G_{\mathbf{a}}$.
Link variables are labeled by line segments, 
$\Sigma_{\mathbf{l}} = \exp(i \pi_{\mathbf{l}})$ that transform
 as bifundamentals under the endpoints of the line
$l=(\mathbf{a},\mathbf{b})$
\begin{eqnarray}
\Sigma_l \rightarrow g_{\mathbf{a}} \Sigma_{\mathbf{l}} g^\dagger_{\mathbf{b}}.  
\end{eqnarray}
Finally the plaquette potentials are interpreted
as shaded in faces and are the product of the link fields that bound
the faces:
$W_{\omega} = \Sigma_{\mathbf{l_1}} \cdots \Sigma_{\mathbf{l_N}}$.
The Lagrangian for a theory space is given by:
\begin{eqnarray}
\LL = \sum_\mathbf{a} \frac{-1}{2 g_\mathbf{a}^2} \Tr F^2_\mathbf{a}
+ \sum_\mathbf{l} \frac{f^2_\mathbf{l}}{4} \Tr 
\big|D^\mu \Sigma_\mathbf{l}\big| ^2
+ \sum_\omega \lambda_\omega f^4 \Tr W_\omega +\hc.
\end{eqnarray}

\begin{figure}[ht]
\centering\epsfig{figure=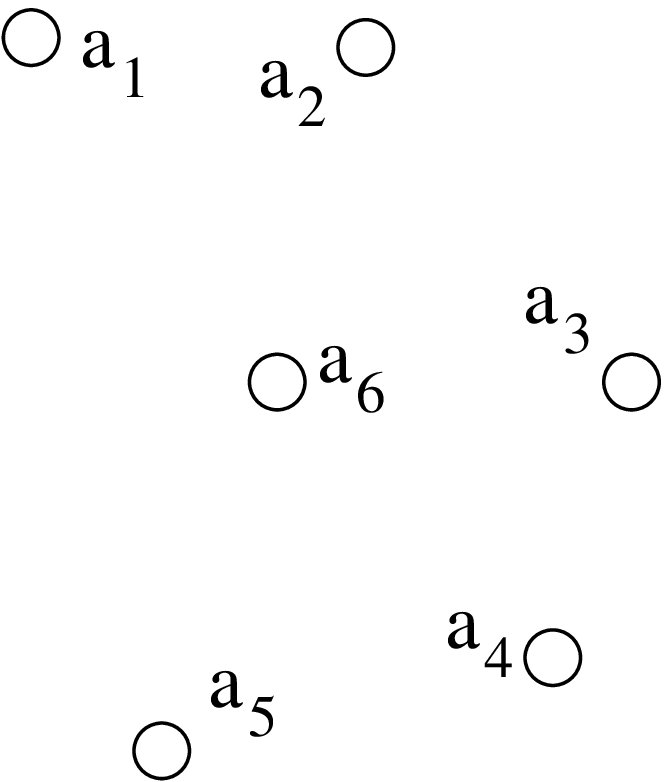, height = 4cm}
\hspace{0.3in}
\centering\epsfig{figure=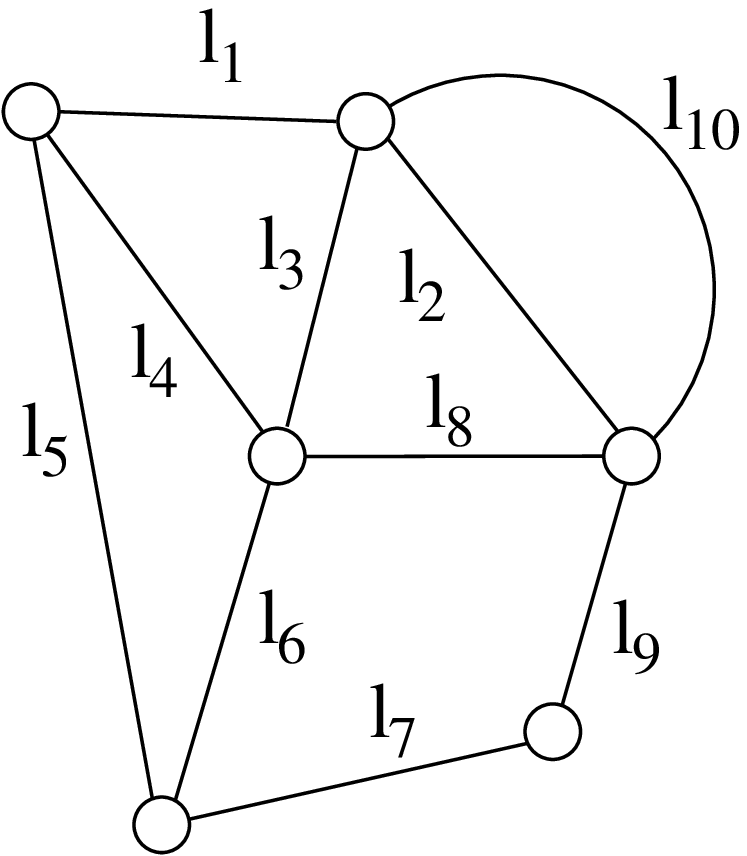,  height = 4.4cm}
\hspace{0.3in}
\centering\epsfig{figure=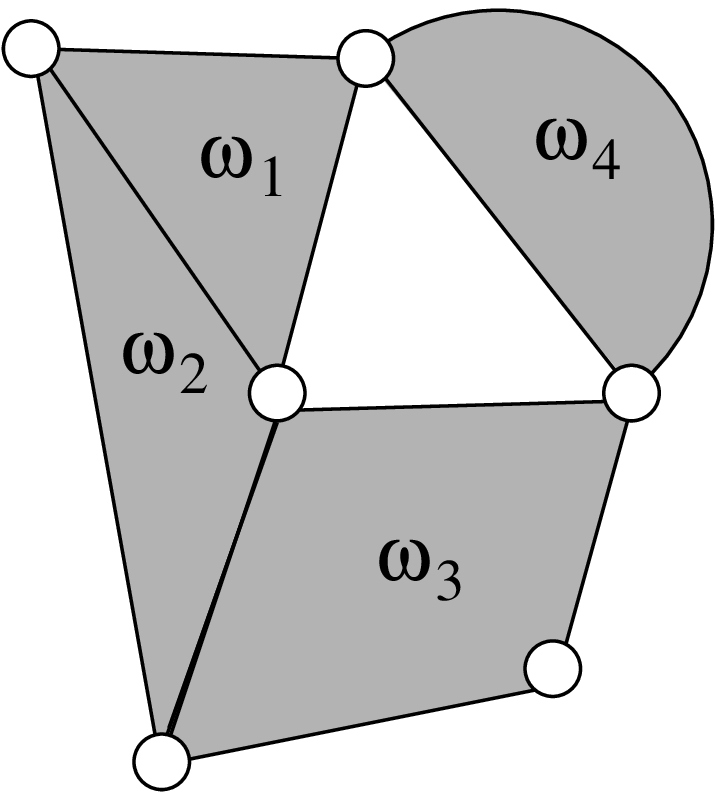, height = 4.1 cm}
\caption{
\label{Fig: TheorySpace}
The geometry of theory space being built up from points, lines,
and faces.  These geometrical objects are identified as
gauge groups, fields, potentials in the action.
}
\end{figure}

The full gauge group of a theory space is given by the product of the gauge
groups associated with each sites: $G_{\text{total}} = \prod_{\mathbf{a}}
G_{\mathbf{a}}$. However, only a small subgroup of this gauge symmetry is
realized linearly on the $\pi_l$. This is the low energy unbroken subgroup
under which:
\begin{equation}
\Sigma_l \rightarrow g \Sigma_l g^{\dagger}
\end{equation}
So long as all the link fields connect two sites, for
each disconnected component of theory space there is an unbroken gauge
symmetry corresponding to the diagonal subgroup of the product of all the
gauge groups associated with the sites in the given component. 

To build realistic models of electroweak symmetry breaking, the Higgs 
must transform as $\mathbf{2_\half}$ under $SU(2)_L \times U(1)_Y$.
 However, if all the gauge groups of a
theory space are the same and the link fields transform as bifundamentals,
the scalars of the theory will be adjoints under the unbroken
gauge group.  One way of solving this problem is
to reduce the gauge symmetry at one of the sites. We will in general take
all the sites to be $SU(3)$ gauge group except one where we will gauge only
$SU(2) \times U(1)$. The link fields are $3\times 3$ matrices and a
link that touch the site of reduced gauge symmetry transform as:
\begin{equation}
\Sigma_l \rightarrow h_{SU(2)}
e^{\frac{i}{6} \theta \mathbf{T_8}} \Sigma_l g_{SU(3)}^{\dagger}
\end{equation}
where $\mathbf{T_8}= \diag(1,1,-2)$\footnote{
The normalization of the $U(1)$ is to have the Higgs doublet have
hypercharge $\half$.
}
and where $h_{SU(2)}$ commutes with $\mathbf{T_8}$.  The unbroken diagonal
subgroup is the electroweak $SU(2)_L \times U(1)_Y$ 
and the scalars of the theory
will decompose into triplets, doublets and singlet of the unbroken
$SU(2)$. The site of reduced symmetry allows for interesting possibilities
that will be discuss in Sec. \ref{Sec: Lifting}, but for the discussion
of the present section, it is irrelevant.

We want to study the low energy physics of these models
at scales beneath the modes that have tree-level masses. 
This can be done by integrating out the massive modes,
but this is a cumbersome procedure.  To integrate out the
heavy modes and have the low energy theory, it is necessary
to find the full spectrum of the theory and to find all  trilinear interactions
involving two light scalars and a heavy scalar and all quartic interactions with only light scalars.  When heavy
scalars are integrated out, trilinear interactions
involving two light scalars and a heavy scalar can exactly cancel
a quartic interaction with only light scalars.  Verifying which
light scalars have a tree-level quartic interaction is therefore rather
intricate and avoiding this procedure is desirable.  
The moduli space captures much of the relevant low energy physics
in the scalar sector and calculating this space will
be the primary goal of this section. We first explain the procedure for
calculating the moduli space of a general theory space and then illustrate
it with several examples.

The moduli space is gauge invariant, meaning that we can
gauge fix in  any convenient manner.  If theory space is arc-wise
connected, then it is possible to draw a simply connected line through 
theory space that touches every point only once.  All the links
along this line can be gauged away and this  procedure completely 
fixes the gauge.  When theory space is not arc-wise connected, there is no
simple rule and we must gauge fix by hand.  
To find the physical spectrum it is more convenient to go to 
unitary gauge which is a more
difficult task. 

After gauge fixing, we minimize the plaquette potential by setting 
the products of
link fields corresponding to faces to the identity matrix.  This
minimization will fix most of the link fields. The interesting part of the
moduli space is then specified by relations between the remaining link
fields.  The flat directions of this moduli space are the little Higgs
of the theory.
To reproduce this moduli space in the low energy
effective action, we include the relations as a
potential so that as we go off the moduli space there is an
energy cost.   Theories that have no relations must have
potentials generated radiatively and therefore have the
same generic problems that typical pseudo-Goldstone bosons
suffer from -- that it is not possible to have a parametric separation
between the cut-off and the vacuum expectation value.
Identifying interesting little Higgs theories reduces to finding
theory spaces with interesting relations.

The procedure of gauge fixing then minimizing the potential is precisely 
equivalent to calculating the fundamental group of theory space 
(or first homotopy group),
see chapter four of \cite{Nakahara:th} for 
more details.  In the equivalence, little Higgs are 
non-contractible cycles on theory space and the low energy potential 
is the relation in the homotopy group.  This links all the relevant
low energy physics to topological properties and is independent of the 
tiling of theory space chosen.  When the tilings are taken to be large,
the physics of theory space is identical to the physics of an extra
dimension.   In the extra dimensional picture, the little Higgs are 
flat gauge connections and are classified by the fundamental group.
In the extra dimensional limit the physics of theory space and of
extra dimensions are identical, however, this equivalence is valid
for any theory space, including ones that bear no resemblance to
an extra dimension.
The relation between the low energy physics and the fundamental group
provides a practical way for both analyzing models as well as 
constructing new models.

\subsection*{Circles and Disks}

A theory space that is topologically a circle is
an example of a theory with a little Higgs.
This theory was analyzed in \cite{Arkani-Hamed:2001nc} and
in more depth in \cite{Hill:2002me}.   
The link fields transform as 
$\Sigma_a \sim \overline{\square}_a\times \square_{a+1}$ and can be written
as exponential: $\Sigma_a = \exp i \pi_a$.  
The Lagrangian is given by:
\begin{eqnarray}
\label{Eq: Lagrangian_circle}
\LL_{S^1} = \sum_a -\frac{1}{2g_a^2} \Tr F^2_a 
+ \sum_a  \frac{f^2_a}{4} \Tr D^\mu \Sigma_a D_\mu \Sigma_a^\dagger  +\cdots
\end{eqnarray}
The ellipses represent higher dimension operators that are
irrelevant at low energies.  
The residual gauge symmetry indicates that there
is a massless gauge boson and $N-1$ massive vector bosons.  Of the 
$N$ non-linear sigma model fields, $N-1$ are eaten by the massive vector 
bosons and one physical massless  scalar is left over. Furthermore, from
Eq. \ref{Eq: Lagrangian_circle}, we see that this scalar does not 
have a tree-level potential because there are no plaquettes.

We will choose to gauge fix in a manner that
eliminates as many of the link fields as possible.
Starting with $\Sigma_1$, we can choose gauge transformations
$g_1$ and $g_2$ so that $\Sigma_1 = \identity$.  Similarly it is possible 
to gauge away $\Sigma_2$ with  $g_3$.  
It is possible to gauge away all but one of the links.
It is not possible to gauge away the last field because 
the last link closes the circle and the gauge freedom
for $g_1$ had already been used to fix $\Sigma_1$.
In this gauge the physical scalar, $\Sigma= \exp(i \sigma)$, 
mixes with the gauge fields, therefore this gauge is inconvenient for 
calculating the physical spectrum of gauge bosons.  
Unitary gauge is more convenient for computing the spectrum
because there is no vector-scalar mixing.
We can interpret $\Sigma$ as a classical modulus of the theory.
This classically massless mode is a pseudo-Goldstone boson
called a little Higgs.   The low energy effective action is just:
\begin{eqnarray}
\LL_{\text{LE}} = -\frac{1}{2 g_D^2} \Tr F^2 
+ \frac{f^2_{\text{LE}}}{4} \Tr\big| D_\mu \Sigma\big|^2 + \cdots
\end{eqnarray}
where $\sigma$ is an adjoint under the unbroken gauge symmetry.
A potential for $\sigma$ that lift the moduli space will be generated at
one loop, however, the only gauge invariant operators are of the 
form $\Tr \Sigma \sim  \cos( \sigma)$.  
The pseudo-Goldstone boson, $\sigma$, can not have significant self-interaction 
without having a significant mass.
This form of the low energy potential is too constrained to be used for 
electroweak symmetry breaking as it does not allow for a parametric
separation between the vacuum expectation value  of the little Higgs 
and the cutoff of the theory.

Next, consider a theory space with the topology of a disk by adding the
plaquette $\Tr \Sigma_1 \cdots \Sigma_N$.  
This space has no non-contractible cycles and therefore has no little Higgs. 
After filling in theory space with more sites, links and plaquettes,
we can make ``holes'' in a disk by omitting plaquettes.  This 
creates non-contractible cycles in theory space.   A theory
space with the topology of a disk with two holes is shown in Fig. \ref{Fig:
Disc2Holes}. We can gauge fix by drawing a line through theory space that 
goes through every points.   Upon minimizing the potential, there are 
two moduli corresponding to the two non-contractible cycles.  These moduli 
are arbitrary non-linear sigma model fields because there is no relation for
the homotopy group.  This means that there is no tree-level potential in
the low energy theory and any deconstruction of this space will be unsuitable 
for electroweak symmetry breaking.   The existence of two little Higgs
does not guarantee a tree level potential.   
Because of the homotopy arguments, a disk with $h$ holes will have
$h$ little Higgs, but none of these scalars will ever have
a tree-level potential because the fundamental group of theory space
is (in the notation of \cite{Nakahara:th})
$\pi_1 = \{C_1, \cdots, C_h: -\}$, where $C_i$ are the non-contractible
cycles on theory space and ``$-$'' represents that there is no
relation between the cycles. 

\begin{figure}[ht]
\centering\epsfig{figure=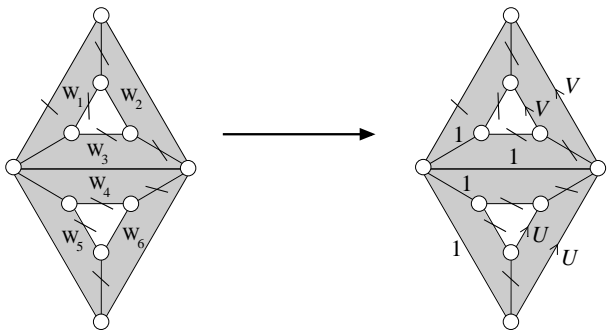, width = 8cm}
\caption{
\label{Fig: Disc2Holes}
A deconstruction of a disk with two holes.
}
\end{figure}

\subsection*{Torus}
\label{Sec: Torus}

A theory space that is topologically a torus has 
two little Higgs.  
The primary new feature with this theory space is the 
appearance of a relation in the definition of the fundamental
group:
\begin{eqnarray}
\label{eq: Homotopy_torus}
\pi_1( T^2) = \{ U, V : U V U^{-1} V^{-1}\}.
\end{eqnarray}
This will lead to a tree-level potential for the little Higgs
associated with the cycles $U$ and $V$.
Consider an $N\times N$ sites deconstruction
of a torus with the sites labeled $(a,b)$.  We will
take our fields to be 
$U_{(a,b)} \sim \overline{\square}_{(a,b)}\times \square_{(a+1,b)}$
and
$V_{(a,b)} \sim \overline{\square}_{(a,b)}\times \square_{(a,b+1)}$.
To make this space topologically a torus, we periodically
identify $(a,b) \equiv (a+N, b) \equiv (a,b+N)$.
This theory breaks the $G^{N^2}$ gauge symmetry down
to the diagonal subgroup $G_D$.  There are $N^2-1$
Nambu-Goldstone bosons that are eaten by the massive
vectors.  From the continuum limit, we suspect that
the potential gives mass to $N^2-1$ of the physical
modes leaving two modes massless. The Lagrangian for theory space is given by:
\begin{eqnarray}
\nonumber
\LL_{T^2} &=& \sum_{a,b} -\frac{1}{g_{(a,b)}^2} \Tr F_{(a,b)}^2
+ \sum_{a,b} 
\frac{f^2_{U (a,b)} }{4} \Tr \big| D_\mu U_{(a,b)}\big|^2
+ \frac{f^2_{V (a,b)} }{4} \Tr \big| D_\mu V_{(a,b)}\big|^2\\
&&
+ \sum_{a,b} \lambda_{(a,b)} f^4 \Tr W_{(a,b)}  + \hc
\end{eqnarray}
where 
\begin{eqnarray}
W_{(a,b)} = U_{(a,b)} V_{(a+1,b)} U^\dagger_{(a,b+1)} V^\dagger_{(a,b)} .
\end{eqnarray}
 We can see that there are two massless modes in this theory from the fact that
the $N^2$ plaquettes terms $W_{(a,b)}$ give masses to $N^2-1$ of the scalars.
\begin{figure}[ht]
\centering\epsfig{figure=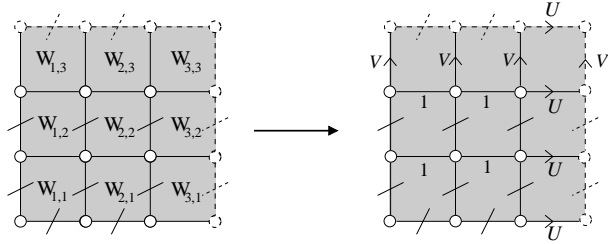, width = 8cm}
\caption{
\label{Fig: Torus}
Gauge fixing of the torus where crossed lines are gauged to
the identity.  The plaquettes are then minimized.  
Plaquette $W_{3,3}$ forces $U V U^{-1} V^{-1} = 1$.
}
\end{figure}

To analyze the model in more details, we first gauge fix to 
eliminate as many fields as possible.  We then minimize
the potentials by requiring that $W_{(a,b)} = \identity$. This procedure is
illustrated in Fig. \ref{Fig: Torus}. We find that the vacuum is given by:
\begin{eqnarray}
U V U^{-1}  V^{-1} = \identity. 
\end{eqnarray}
This is the classical moduli space: two unitary matrices that commute. 
To enforce this in the low energy effective action we 
include this relation as a potential so that there is
an energy cost for going off the moduli space:
\begin{eqnarray}
\LL_\eff = - \frac{1}{2 g_D^2} \Tr F^2
+ \frac{f^2_U}{4} \Tr \big| D_\mu U\big|^2
+ \frac{f^2_V}{4} \Tr \big| D_\mu V\big|^2
+ \lambda_\eff f^4 \Tr U V U^\dagger V^\dagger + \hc .
\end{eqnarray}
There is now a tree-level quartic potential, and masses are induced
radiatively.  This allows a hierarchy between the cut-off and the vacuum
expectation value of little Higgs that will allow
stabilization of the electroweak scale.

\begin{figure}[ht]
\centering\epsfig{figure=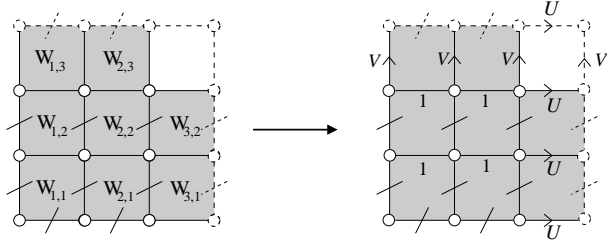, width = 8cm}
\caption{
\label{Fig: TorusHole}
Gauge fixing of the torus where crossed lines are gauged to
the identity.  The plaquettes are then minimized.  
Since plaquette $W_{3,3}$ is absent, there is no relation and the 
moduli space is arbitrary $U$ and $V$ and there is
no low energy potential for the little Higgs.   
}
\end{figure}

If one of the plaquette couplings of the torus is taken to vanish,
the topology of theory space has changed.  In Fig. \ref{Fig: TorusHole}
we compute the fundamental group and find that there
is no relation between the cycles and therefore no low
energy potential for the little Higgs.  We can calculate the
coefficient of the potential for a general torus through a linearized analysis  
by diagonalizing the scalar mass matrix and then integrating
out the massive modes.  We find that the coefficient of
the potential $\lambda_\eff$ is given by:
\begin{eqnarray}
\lambda_\eff^{-1} = \sum_{(a,b)} \lambda_{(a,b)}^{-1}.
\end{eqnarray} 
We see that if any coefficient vanishes, then the low
energy potential vanishes precisely matching the topological
argument.

Toroidal theory spaces of the type shown in
Fig. \ref{Fig: Torus} are not the simplest theory space having the
fundamental group of the torus (Eq. \ref{eq: Homotopy_torus}). Consider a theory space with two sites, four bi-fundamental links
$X_i$ and two plaquettes:
\begin{eqnarray}
\label{Eq: Nelson-Katz}
V(X) = - \lambda_1 f^4 \Tr X_1 X_2^\dagger X_3 X_4^\dagger
 - \lambda_2 f^4 \Tr X_2 X_3^\dagger X_4 X_1^\dagger.
\end{eqnarray}
This theory was first analyzed in \cite{2sites}. 
It can be easily analyzed by first
gauge fixing $X_1$ to $\identity$ and then 
solving for $X_4 = X_2^\dagger X_3$.  We are
left with the relation
\begin{eqnarray}
 X_2 X_3^\dagger X_2^\dagger X_3 = \identity 
\end{eqnarray}
which is the commutator potential of a torus. 
One can show that this theory space is related
to the $2\times 2$ torus by orbifolding by a translational symmetry
that sends all points $(i,j) \rightarrow (i+1, j+1)$.
This symmetry acts freely and does not change the homotopy of the
space and therefore does not change the 
little Higgs or their self-interaction. The physics of this theory space
is studied in more details in \cite{2sites}.

\subsection{Reverse Engineering}
\label{subsec: Reverse}

Finding the low energy physics from a theory space is a straight-forward
procedure of gauge fixing then minimizing the potential.  There is also
an intuitive procedure for taking a low energy potential in the form of a
product of nonlinear sigma model fields  and finding a high energy theory 
that produces it at low energy.
This construction is reverse engineering the theory space from the
low energy potential.  
The most interesting theories to consider are the minimal ones.   
It is not difficult to conclude that the simplest potential that 
is viable for electroweak symmetry breaking is
$\Tr U V U^\dagger V^\dagger$.  This means that the theory
space that produces this potential is homotopically equivalent
to the torus.  The
simplest  such theory with more than one site is the two sites four links model of the previous
section.  To illustrate this construction we will use non-minimal
models that are still viable models of electroweak symmetry breaking.

\begin{figure}[ht]
\centering\epsfig{figure=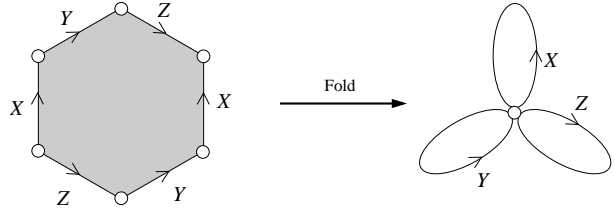, width = 8cm}
\caption{
\label{Fig: XYZSkel}
The minimal theory space with three cycles, $X$, $Y$,
$Z$ and the potential $V = \Tr X Y Z X^{-1} Y^{-1}Z^{-1}$.
The arrows along the links indicate whether the fields are
$X_i$ or $X_i^{-1}$.
}
\end{figure}

\begin{figure}[ht]
\centering\epsfig{figure=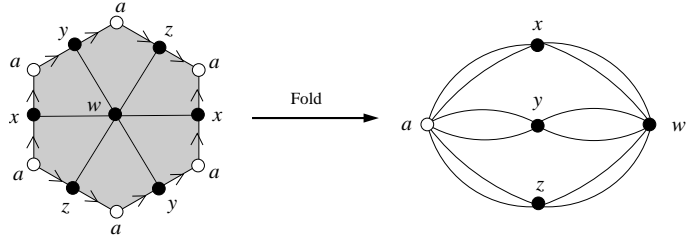, width = 9cm}
\caption{
\label{Fig: XYZModel}
A larger deconstruction of the $XYZX^{-1} Y^{-1}Z^{-1}$
three cycle model where we have introduced 
four new gauge groups ($x$, $y$, $z$, $w$).
The plaquette structure in the unfolded deconstruction
is obvious, but the field content is harder to visualize
because of the identifications.  In the folded version,
the gauge and field content is clear, but the plaquettes structure is obscured.
}
\end{figure}

\begin{figure}[ht]
\centering\epsfig{figure=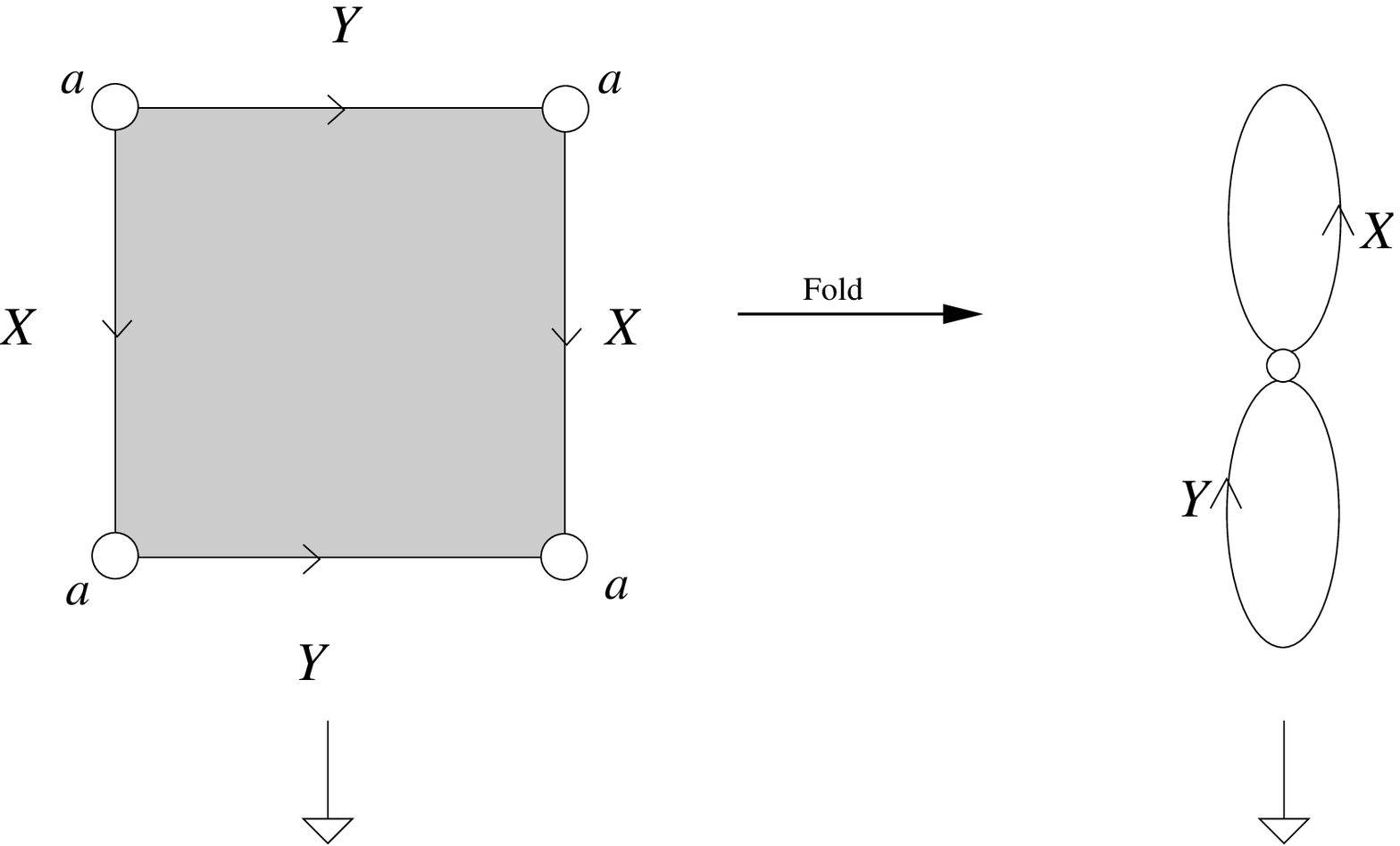, width = 6.5cm} \\
\centering\epsfig{figure=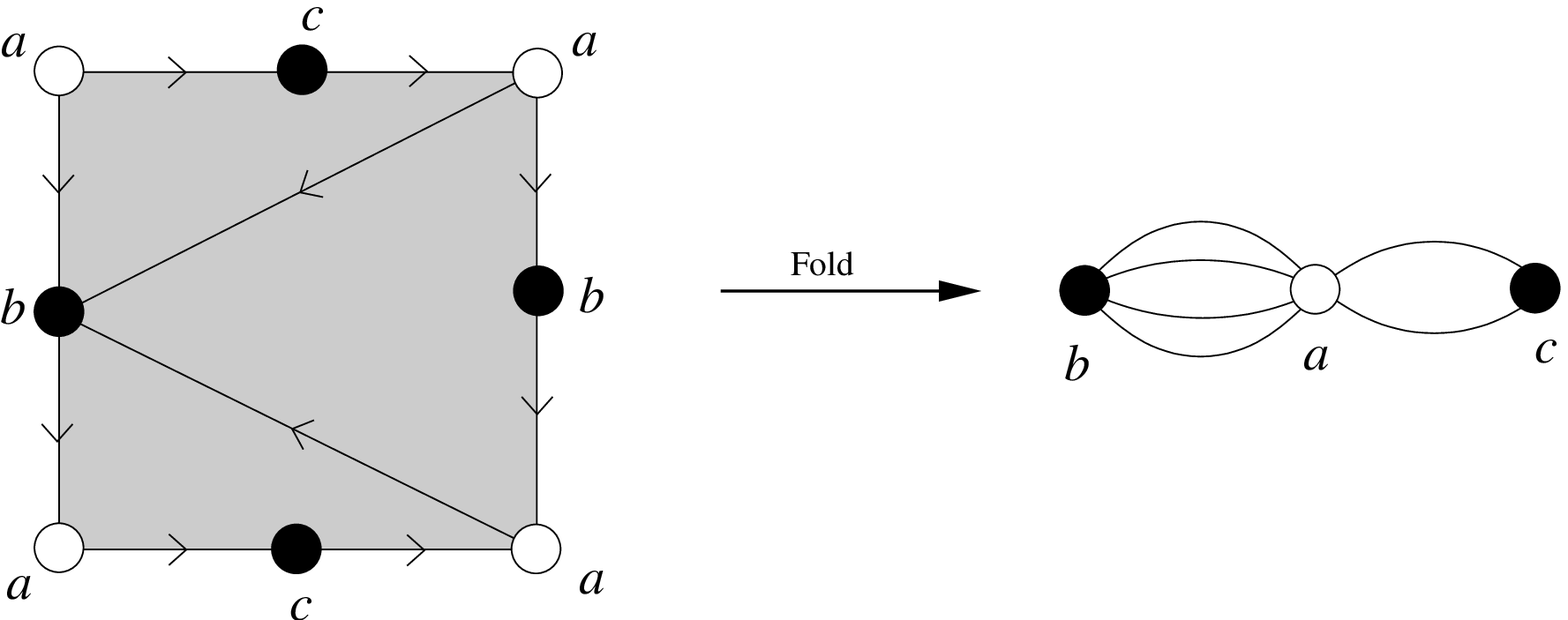, width = 7cm}\\
\centering\epsfig{figure=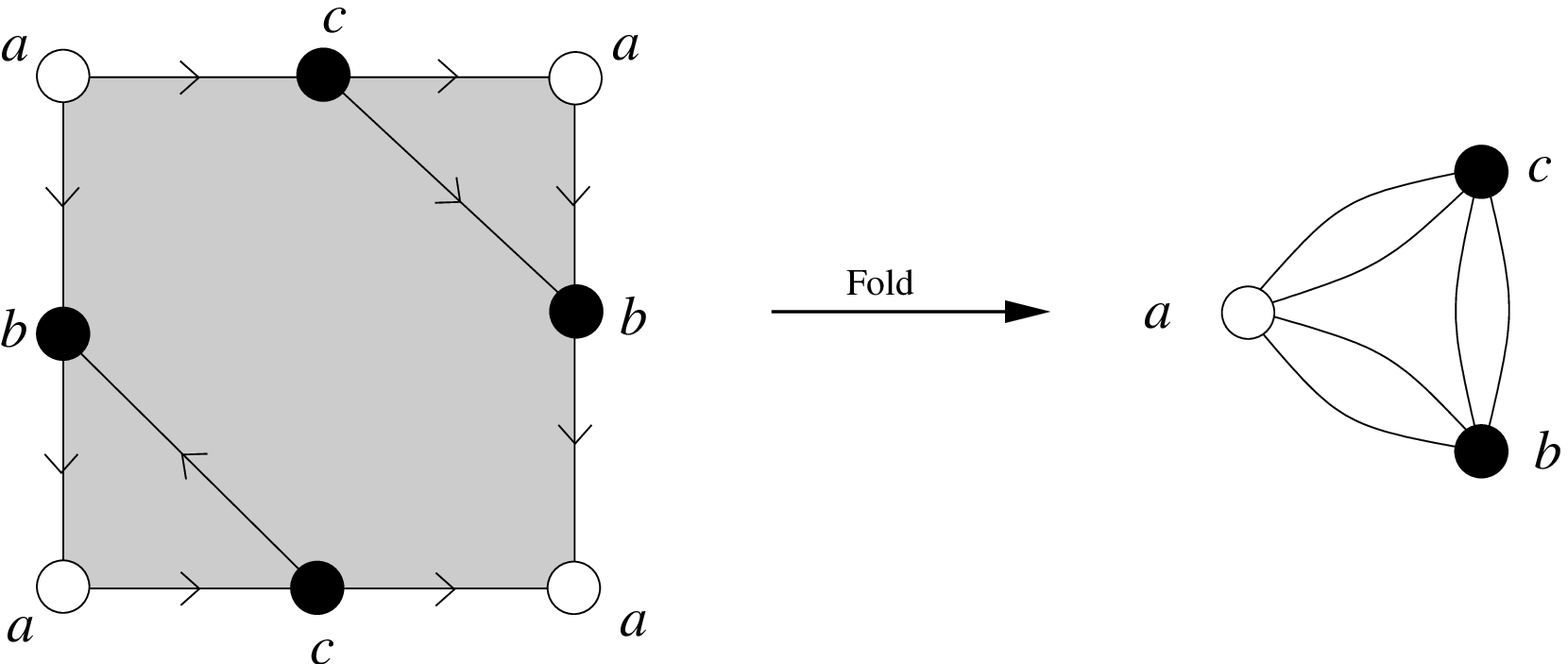, width = 6.5cm}\\
\centering\epsfig{figure=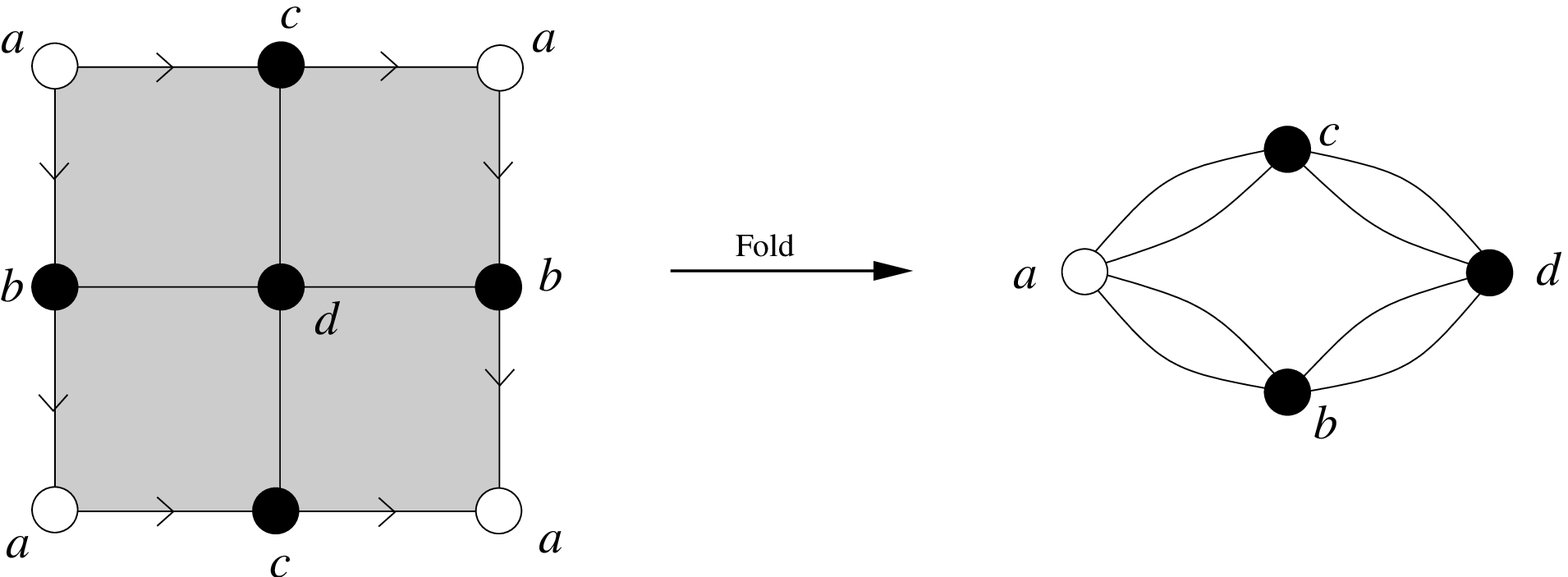, width = 7.5cm}
\caption{
\label{Fig: Torus2}
Alternative deconstructions of a torus. The last figure is the $2\times 2$
torus of \cite{Arkani-Hamed:2002pa}.
}
\end{figure}

Given a set of non-linear sigma model light fields $X_i$ and a potential
$V(X_i)$ that is a product of the fields and their
inverses, we draw out the potential as a polygon
with each side being the corresponding link field. 
Each link begins and ends at the same site, $a$.
For instance consider three light fields $X$, $Y$, $Z$
and a potential $V = \Tr X Y Z X^{-1} Y^{-1}Z^{-1}$.
In Fig. \ref{Fig: XYZSkel}, we draw out the unfolded
and folded versions of this theory space.
Any theory space that tiles this minimal 
version of theory space will have the same low
energy potential.  Dividing
the plaquettes and links by placing new points and
links in theory space will not change the low energy
potential.  
For instance we can divide
the theory space in Fig. \ref{Fig: XYZSkel}
up in Fig. \ref{Fig: XYZModel}. We can also build different theory
spaces that have the same low energy physics as the torus.
Figure \ref{Fig: Torus2} shows three such theory spaces. 
They are obtained by requiring a low
energy potential of the form $X Y X^{-1} Y^{-1}$ and tiling the original
construction in different manners.

\begin{figure}[ht]
\centering\epsfig{figure=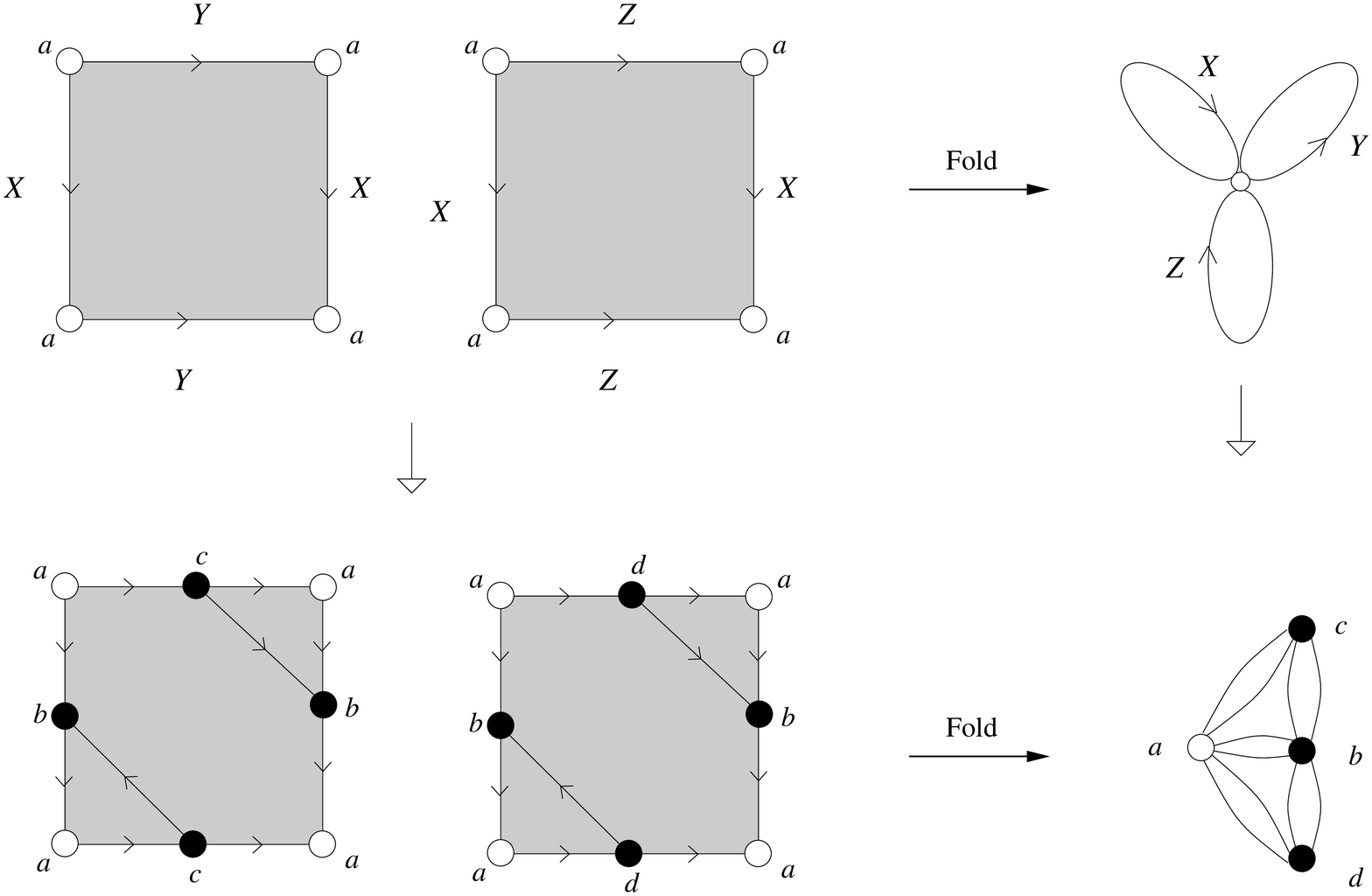, width = 9cm} 
\caption{
\label{Fig: Two Relations}
Construction of a theory space with relations $X Y X^{-1} Y^{-1}$
\emph{and} $ X Z X^{-1} Z^{-1}$. Note that the two squares must be
tiled with different links.}
\end{figure}

Finally, some spaces have fundamental groups with more than one relation.  
To construct theory spaces that are homotopically equivalent
to these spaces we draw the multiple relations as disjoint diagrams 
although theory space is connected.  
In Fig. \ref{Fig: Two Relations} a theory space with a fundamental
group
\begin{eqnarray}
\pi_1 =  \{ X, Y, Z : X Y X^{-1} Y^{-1} , X Z X^{-1} Z^{-1}\}
\end{eqnarray}
is constructed.

We have shown how to analyze and build theory spaces with classically
massless Higgs and order one quartic interactions. This is not
sufficient to ensure that a theory space can be used for electroweak
symmetry breaking, as radiative
corrections might make the Higgs too heavy. We will show in the next
section that in order for that not to be the case, theory spaces must
satisfy mild constraints but there is still an arbitrariness to the
theory spaces that produce a given low energy physics.

\section{Radiative Corrections}
\label{Sec: Radiative Corrections}
 
Without gauge couplings and plaquette interactions, a theory
space with $M$ link fields has a $G^{2 M}$ global chiral symmetry, under
which each link field transform as bifundamental under 
independent global symmetries:
\begin{equation}
\Sigma_l \rightarrow L_l\Sigma_l R_l^\dagger
\end{equation}
Without couplings,
the link fields are exact Goldstone bosons with only derivative
interactions.  Once gauge and plaquette couplings are 
included, some set of the chiral symmetries are broken. 
The coupling constants may be viewed as spurions that 
give rise to masses and non-derivative interactions.  
The essential feature of little Higgs theories that guarantees
ultraviolet insensitivity is that generation of operators
containing mass terms for the little Higgs requires 
many spurions.
Consequently, since ultraviolet physics is analytic in the
parameters, quadratically divergent contributions to
the little Higgs mass are suppressed by many loop factors. 

When building theory spaces there must be enough
spurions so that there are no one loop quadratic divergences.
However, even if the one loop quadratic divergences are absent, 
generically there will be a one loop finite contribution to the little
Higgs mass so long as the little Higgs is not an exact Goldstone boson.
Infrared physics is not analytic in the parameters and
the finite contribution is of the order of:
\begin{eqnarray}
m_{\text{LH}}^2 \sim  \frac{g^2}{(4 \pi)^2} M_{\text{H}}^2
\end{eqnarray}
where $M_H$ is the mass of lightest new state which generically
is of order $M_H\sim g f$.  Using this relation we find:
\begin{eqnarray}
m_{\text{LH}}^2 \sim  \frac{g^2}{(4 \pi)^2} g^2 f^2 \sim 
\frac{g^4}{(4 \pi)^4} \Lambda^2
\end{eqnarray}
where $\Lambda \sim 4 \pi f$ is the ultraviolet cut-off of the
theory.  The infrared contributions are of the same
order of magnitude as a two loop quadratic divergence.
Therefore, it is unnecessary to eliminate anything but the one loop
quadratic divergence.  The only benefit of eliminating 
divergent contribution of higher loop order would
be that the little Higgs mass would be calculable because the mass 
would be dominated by infrared physics, as opposed to having ultraviolet
and infrared physics providing parametrically the same contribution.
Another possible reason for eliminating more than one loop
quadratic divergences would be if a coupling was so strong
so that loops involving this coupling  were not suppressed.

Having to only eliminate the one loop quadratic divergences,
the constraints on theory space are very mild and can be stated
simply:
\begin{description}
\item[Gauge Sector:] Every link must connect two different sites.
\item[Scalar Sector:] No plaquette can contain the same link twice.
\end{description}
We can prove these rules by computing the quadratically 
divergent part of the one loop Coleman-Weinberg potential. 
We turn on a little Higgs
background fields and calculate $\Tr M^{\dagger}M$ where $M$
is the mass matrix of the theory in the presence of the background.

We first consider the gauge sector and show that gauge interactions never
produce one loop quadratic divergences so long as all the link connect
two different sites or equivalently all link fields are in bifundamentals
as opposed to adjoint representations.
Consider a link field between two different sites $i$ and $j$.  The gauge
boson mass matrix comes from the covariant derivative,
$A_i^a  M^2{}^{ij}_{ab}[\tilde{U}] A_j^b$, where $a,b$ are gauge indices and
\begin{eqnarray*}
M^2_{ab}[\tilde{U}] = \frac{f^2}{4} 
\left( \begin{array}{cc} 
\half g_i^2 \delta_{ab}& -g_i g_j m_{ab}[\tilde{U}] \\
-g_i g_j  m_{ab}^\dagger[\tilde{U}]
& \half g_j^2 \delta_{ab}
\end{array}\right) \hspace{0.6in}
m_{ab}[\tilde{U}]=\Tr \mathbf{T}_a \tilde{U} \mathbf{T}_b \tilde{U}^\dagger
\end{eqnarray*}
The important point is that $\diag M^2$ is always independent of
the background field, $\tilde{U}$, 
and therefore will never produce a one loop quadratic divergence
for any link field mass.  If a field is in the adjoint, then
this argument will break down and  a one loop quadratic divergence
will appear.

We now turn to the scalar sector.  Consider an arbitrary plaquette:
\begin{eqnarray}
V(U_i) = - \lambda f^4 \Tr M_1 U_1 \cdots M_N U_N + \hc
\end{eqnarray}
where $M_i$ are arbitrary matrices.
We rewrite the link fields 
as a linearized fluctuations, $u_i$, and a background fields, $\tilde{U}_i$:
\mbox{$U_i = \exp(i u_i) \tilde{U}_i$.} By dividing $U_i$ in this way, the
background field drops off the kinetic term and we can extract the mass of
$u_i$ directly from the potential without having to worry about putting
the kinetic term in canonical form. We expand
out the plaquette to quadratic order in the
fluctuations and find the mass matrix,
$u_i^a M^2{}_{ab}^{ij} u_j^b$.   The diagonal of the
mass matrix is
\begin{eqnarray}
\diag M^2{}_{ab}^{ij} 
\sim \lambda f^2 \Tr M_1 \tilde{U}_1 \cdots M_i \mathbf{T}_a \mathbf{T}_a 
\tilde{U}_i 
M_{i+1} \tilde{U}_{i+1} \cdots M_{N} \tilde{U}_{N}
\end{eqnarray}
 where $\mathbf{T}_a$ are gauge group generators. 
Then summing over the diagonal
entries of the mass matrix  and using $\sum_a \mathbf{T}_a ^2 \sim
\identity$, we find $
\Tr M^2 \propto M_1 \tilde{U}_1 \cdots M_i \tilde{U}_i M_{i+1}
\tilde{U}_{i+1} \cdots M_{N}\tilde{U}_{N}$ which is just the plaquette
operator. Since, by definition, the plaquettes do no contain mass term for
the little Higgs, this shows that plaquettes never produce one loop quadratic divergences
to the little Higgs mass unless fields appear in plaquettes
more than once.
If a field appears in a plaquette more than once, than this argument
will break down because the mass matrix will have a more
complicated form with $\tilde{U}$ dependent diagonal entries.

We are left with two requirements for a theory space to 
have no one loop quadratic divergences: that
no link begins and ends on the same point -- that 
no link field is in an adjoint representation,
and that no plaquette contains a link twice.
These constraints can be easily satisfied, even with small
theory spaces. 
These requirements place restrictions on the minimal field content 
at the TeV scale.  For instance, there must be at least a second
$SU(2)\times U(1)$ (or $SU(3)$) gauge symmetry broken 
around the TeV scale with massive gauge bosons $W'$ and $B'$.
There must also be a massive multiplet of triplet, doublet and singlet
scalars $\phi$, $h$, and $\eta$
at the TeV scale to ensure that the scalar potential does not induce
a one loop quadratic divergence.

\section{Fermions}
\label{Sec: Fermions}

The Standard Model Higgs is a pseudo Goldstone boson in little 
Higgs models and has
the same quantum numbers as the kaon.   The Higgs mass is
only protected from one loop quadratic divergences
if we preserve some of the global $SU(3)$ chiral symmetry.
In the gauge and scalar sectors of these theories this was
automatic at one loop, however, in the fermion sector
one loop quadratic divergences are possible
if all the $SU(3)$ chiral symmetry that is protecting the Higgs 
mass is broken by one coupling.

It is useful to write the Standard Model fermions as incomplete $SU(3)$
triplets at the $SU(2)\times U(1)$ site $\mathbf{0}$ in order to
make manifest the $SU(3)$ symmetries we want to preserve in
the Yukawa couplings.
\begin{eqnarray}
Q = \left( \begin{array}{c} q\\0\end{array} \right)
\hspace{0.2in}
U^c = \left( \begin{array}{c} 0\\0\\u^c\end{array} \right)
\hspace{0.2in}
D^c = \left( \begin{array}{c} 0\\0\\d^c\end{array} \right)
\hspace{0.2in}
L = \left( \begin{array}{c} l\\0\end{array} \right)
\hspace{0.2in}
E^c = \left( \begin{array}{c} 0\\0\\e^c\end{array} \right)
\end{eqnarray}
Under a $U(1)_{\mathbf{0}}$ transformation, $\theta$, these fields
transform as:
\begin{eqnarray}
Q \rightarrow e^{ \frac{i}{6} \theta} Q
\hspace{0.2in}
U^c \rightarrow e^{\!-\frac{2i}{3} \theta} U^c
\hspace{0.2in}
D^c \rightarrow e^{\frac{i}{3} \theta} D^c
\hspace{0.2in}
L \rightarrow e^{\!-\frac{i}{2} \theta} L
\hspace{0.2in}
E^c \rightarrow e^{i \theta} E^c
\end{eqnarray}
At low energies, the effective coupling to the little Higgs is
just through a Wilson line operator $W$ that stretches
from site $\mathbf{0}$ back to site $\mathbf{0}$.
Under a $U(1)_{\mathbf{0}}$ transformation, $W$ transforms
as:
\begin{eqnarray}
W \rightarrow  
\exp\Big( \frac{i}{6} \mathbf{T_8} \theta \Big)
\;W \;
\exp\Big(\!-\!\frac{i}{6} \mathbf{T_8} \theta\Big)
\end{eqnarray}
with $\mathbf{T_8} = \diag (1,1,-2)$.  Let us introduce
projections matrices $P_1 = \diag (1,1,0)$ and $P_2 = \diag(0,0,1)$.
The gauge invariant Yukawa couplings\footnote{
The low energy $ll hh$ dimension five Yukawa coupling that gives a neutrino
mass is written in this language as $L^T P_1 W^T P_2 W P_1 L$.} 
are given by:
\begin{eqnarray}
y_u f Q^T P_1 W P_2 U^c 
\hspace{0.6in}
y_d f Q^T P_1 W^* P_2 D^c
\hspace{0.6in}
y_e f L^T P_1 W^* P_2 E^c .
\end{eqnarray}
These couplings arise from an ultraviolet completion in the $\sim 10$ TeV
range.  Having particles that carry flavour at this scale can produce
unacceptably large flavour changing neutral currents
\cite{Chivukula:2002ww}.  Flavour physics places constraints
on possible completions.  A simple solution is to complete these theories
into supersymmetric linear sigma models at this scale.
These couplings introduce quadratic divergences of the form:
\begin{eqnarray}
\LL_{\eff} = \frac{y_f^2}{16 \pi^2} f^2 \Lambda^2 \Tr P_1 W P_2 W^\dagger .
\end{eqnarray}
This is just the usual quadratic divergence to the Higgs mass coming from
Yukawa couplings.
For everything, but the top quark, the Yukawa couplings are small enough
so that the quadratic divergences are small enough to be ignored.
For the top quark, removing the one-loop quadratic divergence is of
paramount importance.  A solution was discussed in 
\cite{Arkani-Hamed:2001nc, Arkani-Hamed:2002pa} where additional
Dirac fermions were introduced on intermediate $SU(3)$ sites.
The key ingredient was preserving at least one of the $SU(3)$ global
symmetries protecting the Higgs mass.
In this note we will consider an alternative mechanism.
We can imagine introducing an Dirac $SU(2)$ doublet $S$, $S^c$  
such that we complete $U^c$ into an $SU(3)$ triplet:
\begin{eqnarray}
U^c  = \left( \begin{array}{c} S^c\\u^c\end{array} \right). 
\end{eqnarray}
With the Lagrangian:
\begin{eqnarray}
\label{Eq: Top Div}
\LL_{\text{top}} =  y_u f Q^T P_1 W U^c + m_S S S^c +\hc
\end{eqnarray}
the one-loop quadratic divergence is $\Tr P_1 W W^\dagger$, where
there is not a second projection matrix because of the global
$SU(3)$ symmetry of $U^c$.   If $W$ is unitary, i.e. a product
of link fields, then this removes the one loop quadratic divergence.
The chiral symmetry  protects the little Higgs' mass.
Similarly to the gauge and scalar sectors, we now have a rule for avoiding
quadratic divergence in the fermionic sector:
\begin{description}
\item[Fermion Sector:] The top Yukawa couplings must preserve either
a left ($W \rightarrow L W$)  or right ($W\rightarrow W R^\dagger$)
chiral symmetry.
\end{description}
The effective top quark Yukawa coupling is
\begin{eqnarray}
y_\eff^{-2} =  (y_u)^{-2} + (m_S/f)^{-2}
\end{eqnarray}
meaning that $m_S/f$ and $y_u$ should both be at least order unity
to have an adequately large top Yukawa coupling.

\subsection{Plaquettes from Yukawa Interactions}
\label{Sec: Yuk2Plaq}

We now restrict ourselves to the model of \cite{2sites}
involving two sites and four links.   If we consider
an alternate Wilson line: $W_1 = X_1 X_2^\dagger + X_4 X_3^\dagger$,
for the top quark then we find a quadratic divergence in 
Eq. \ref{Eq: Top Div}.  However, this divergence is to
the operator $\Tr P_1 X_1 X_2^\dagger X_3 X_4^\dagger$ --
 one of the requisite plaquettes.  This indicates that
with this choice of Yukawa coupling, it is \begin{it} unnatural\end{it}
for the coefficient of this plaquette to be \begin{it}small\end{it}.
In other words, if we choose to set the tree-level coefficient of
the operator to zero, it will be generated at one-loop with an
order $f^4$ coefficient,  precisely the value we want.  
This is only one of the plaquettes in the model of \cite{2sites},
 but with a slightly more elaborate fermion sector it is
possible to generate both plaquettes from the top sector alone.
The emphasis is that plaquette operators 
are naturally generated with a sizeable coefficients from
physics below 10 TeV.

A simple realization of top physics inducing the entire
Higgs potential uses an additional colored weak
doublet Dirac fermion $\tilde{q}$, $\tilde{q}^c$.
Introducing two Wilson lines $W_1 = c_1 X_1 X_2^\dagger + c_1' X_4 X_3^\dagger$
and $W_2 = c_2 X_4 X_1^\dagger + c_2' X_3 X_2^\dagger$, we couple one
Wilson line to each two quark doublet 
in the Yukawa interactions:
\begin{eqnarray}
\LL_{\text{top}} = y_u f Q W_1 U^c + \tilde{y}_u f \tilde{Q} W_2 M U^c
+ m_{\tilde{q}} \tilde{q} \tilde{q}^c  + m_S S S^c .
\end{eqnarray}
where $M = \diag(1,1,i)$ is a unitary matrix of phases and
$\tilde{Q}= (\tilde{q},0)$.
The one loop quadratic divergence gives each plaquette:
\begin{eqnarray}
\LL_\eff = 
\frac{y_u^2}{16 \pi^2} f^2 \Lambda^2 \Tr P_1 |W_1|^2
+ \frac{\tilde{y}_u^2}{16 \pi^2} f^2 \Lambda^2 \Tr P_1 |W_2|^2.
\end{eqnarray} 
The one loop Coleman-Weinberg analysis gives a negative contribution
to the Higgs mass driving it negative and breaking electroweak symmetry.

\section{Lifting States}
\label{Sec: Lifting}

As mentioned in Sec. \ref{Sec: Homotopy}, the theory spaces we are
considering for realistic models have mostly $SU(3)$ sites, one
$SU(2) \times U(1)$ site, and $3 \times 3$ matrix link fields transforming
as bifundamentals under the $SU(3)$ chiral symmetries. Scalars 
decompose into triplets, doublets and singlets under the unbroken $SU(2)$. 
In all the models presented in the previous sections,
the triplets, doublets and singlets were classically degenerate. 
To construct realistic theories we need light doublets, but 
the triplets and singlets appear as extra adjoint matter that
appears to make the doublets into a full $SU(3)$ adjoint multiplets.
Finding the minimal $100 \text{ GeV}$ field content is an interesting
question for phenomenological signatures of the model.
The natural question is whether it is possible to remove the light
triplets and singlets from the $100 \text{ GeV}$ spectrum.
Until now we have considered plaquettes that were $SU(3)$ symmetric,
and treated triplets, doublets and singlets on equal footing. 
In this section we generalize plaquette operators to include
matrices that are invariant under the $SU(2) \times U(1)$ gauge symmetry,
but break the $SU(3)$ chiral
symmetry. This will allow lifting the extra adjoint matter up
to the TeV scale while leaving the doublets at the $100 \text{ GeV}$ scale.

The new types of operators that we will consider are of the form:
\begin{equation}
\Tr M \Sigma_{\mathbf{0},\mathbf{n}} \Sigma_{\mathbf{n},\mathbf{m}} \cdots
\end{equation}
with $\mathbf{0}$ being the $SU(2)\times U(1)$ site and 
$M= \text{diag}(a,a,b).$
The analysis of the low energy physics of theory spaces that contain these
generalized plaquettes proceeds as before,  by first gauge fixing and then
minimizing the potential plaquette by plaquette. However, the plaquette
might not be minimized when the product of link fields is the identity as
before. For a plaquette of the form:
\begin{eqnarray}
\label{Eq: Phases Potential}
 - \lambda \Tr M \Sigma +\hc
\end{eqnarray}
with $\lambda$ real and positive,  there are three 
different phases for the minimum, depending on the choice of $a$ and $b$.
\begin{figure}[ht]
\centering\epsfig{figure=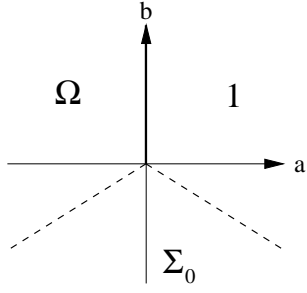, width = 4cm}
\caption{
\label{Fig: Phases}
Minima of the potential in Eq. \ref{Eq: Phases Potential}
labeled in Eq. \ref{Eq: Phases}.  
}
\end{figure}
\begin{eqnarray}
\label{Eq: Phases}
\begin{array}{ccc}
a>0 & b >-\half |a| &
\langle \Sigma \rangle = \identity\\
a<0 & b>- \half |a| & 
\langle \Sigma \rangle = \Omega \\
&b < - \half|a|&
\langle \Sigma \rangle =  \Sigma_0
\end{array}
\end{eqnarray}
with 
$\Sigma_0 = \exp\big( i \mathbf{T_8} \eta_0\big)$,
$\eta_0 = \cos^{-1}( -2b/a)$,
and 
$\Omega = \diag( -1 -1,1)$,
$\mathbf{T_8} = \diag( 1,1,-2)$.
Typically the $\Sigma_0$ vacuum is uninteresting because
it produces tree level masses for all the fields and we will not consider
it any further. 
 
The resulting moduli space might not be $SU(3)$ symmetric, and when the
link fields are expanded around the appropriate vacuum, the number of
triplet and singlet zero modes might be different than the number of
doublet zero modes. To see how this happens, consider a general $3\times 3$ special unitary matrix:
\begin{equation}
Z = \exp (i z) = \exp i \begin{pmatrix} \phi+\eta&
h\\h^{\dagger}&-2 \eta \end{pmatrix}
\end{equation}
then
\begin{eqnarray}
\label{Eq: Omega}
\Omega Z \Omega = \exp(i \Omega z \Omega)= \exp i\begin{pmatrix}
\phi + \eta & -h \\ -h^{\dagger} & -2 \eta \end{pmatrix}
\end{eqnarray}
and a relation of the form
\begin{eqnarray}
\Omega Z \Omega Z = \identity  
\hspace{0.35in}
\Rightarrow
\hspace{0.35in}
V = - \lambda f^4 \Tr Z \Omega Z \Omega \sim
\lambda f^2 \Tr\big( \phi^2 + \eta^2\big) + \cdots
\end{eqnarray}
indicates that around $Z=\identity$, the triplet and singlet, $\phi$ and
$\eta$ are massive while the doublet $h$ is massless. We can now use this tool to lift the triplet and singlet zero modes
that were present in the models considered until now. The most obvious set
of relations that would produce this result is given by:
\begin{eqnarray}
\label{Eq: Relations}
UVU^{-1}V^{-1} = \identity 
\hspace{0.3in}
U\Omega U \Omega = \identity
\hspace{0.3in}
V \Omega V \Omega = \identity
\end{eqnarray}
The first relation guarantees the presence of a commutator quartic
potential as in the torus, and the last two relations, when expanded around
$U,V = \identity$ lift the singlet and triplet zero modes.

\begin{figure}
\centering\epsfig{figure=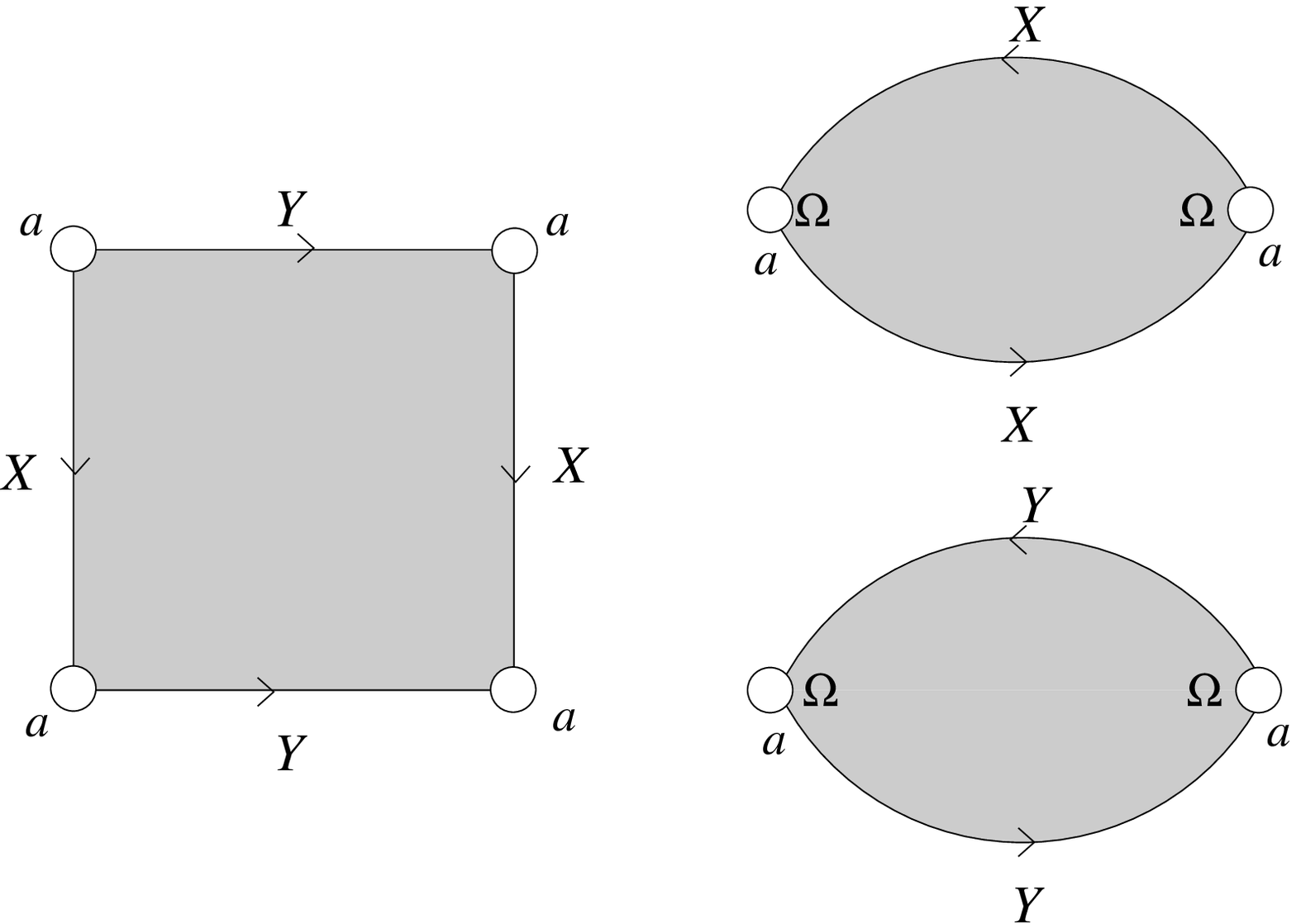, width = 6cm}
\hspace{0.5in}
\centering\epsfig{figure=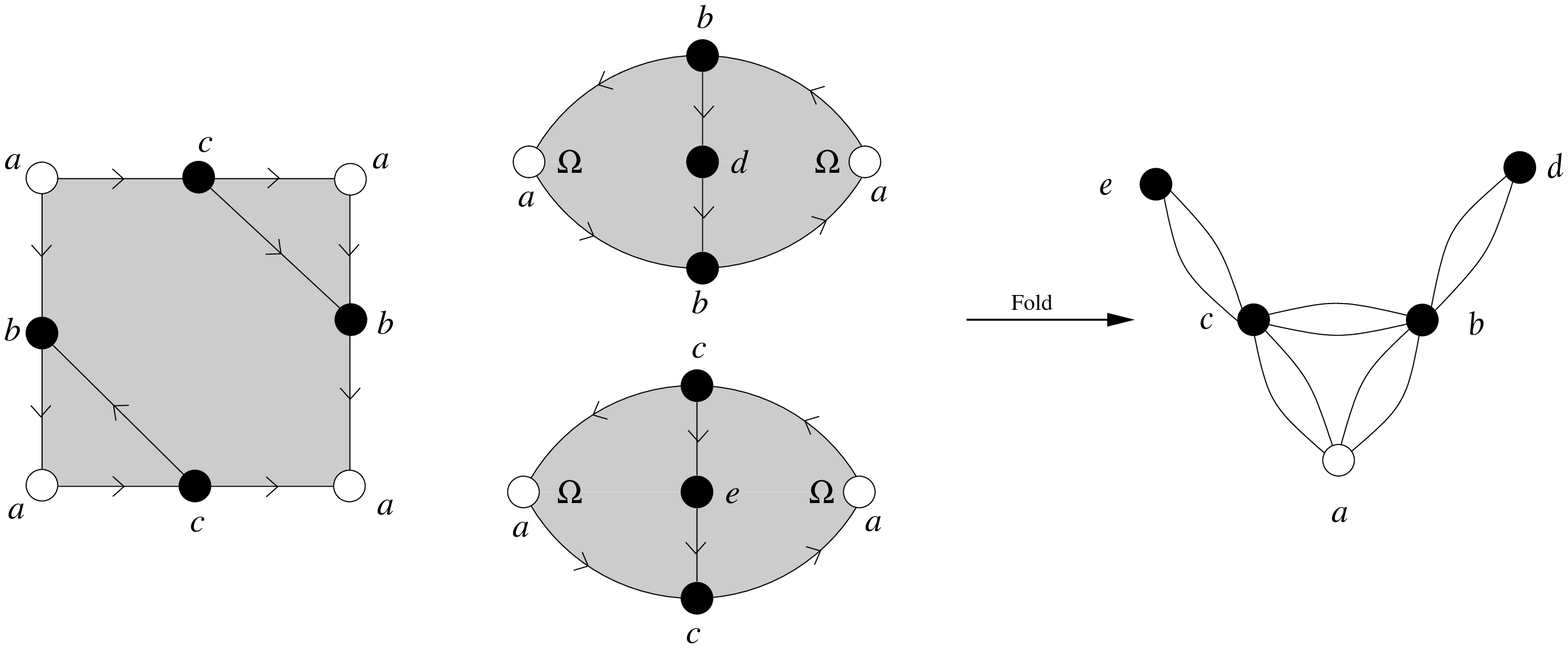, width = 10cm}
\caption{
Construction of a theory space with relation in Eq. \ref{Eq: Relations}
that lift triplet and singlet Higgs. The starting point is the first
picture where the three relations are drawn with the $\Omega$ inserted. The
second picture shows a tiling that has no one loop quadratic 
divergences.
\label{Fig: moose}
} 
\end{figure}

We now need to build a theory space which yield those relations. We use a
very similar procedure to the one described in Sec. \ref{subsec: Reverse}. 
As before, we first draw the relations using only one site but we now 
insert $\Omega$ as they appear in the relations.  We then tile this 
construction in a way that satisfies the rules mentioned earlier. 
The insertion of $\Omega$ represents a plaquette that is minimized at 
$\Omega$.  Fig. \ref{Fig: moose} shows the building of the theory space in
question.

\subsection{Minimal model}

\begin{figure}[ht]
\centering\epsfig{figure=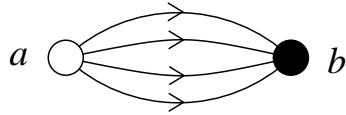, width = 4.5cm}
\caption{
\label{Fig: Minimal}
A minimal model for electroweak symmetry breaking
by little Higgs.
}
\end{figure}

We can also build simpler theory spaces with the same relations. Consider the two sites model presented in
Sec. \ref{Sec: Torus}. In addition to the plaquettes in
Eq. \ref{Eq: Nelson-Katz} we add two new plaquettes containing 
$\Omega$. The total
potential is given by:
\begin{eqnarray}
\nonumber
V &=& -\lambda_1  X_1 X_2^\dagger  X_3 X_4^\dagger
-\lambda_2 \Tr X_1^\dagger  X_2 X_3^\dagger X_4\\
&&-\lambda_3 \Tr \Omega X_1 X_2^\dagger \Omega X_4 X_3^\dagger
-\lambda_4 \Tr \Omega X_1 X_4^\dagger \Omega X_2 X_3^\dagger
\end{eqnarray}
The analysis of this model is straight forward.
We can gauge fix by setting $X_1 = \identity$. We then minimize the first
plaquette which gives $X_3 = X_2 X_4$. Minimization of the second plaquette
gives $X_2 X_4^\dagger X_2^\dagger X_4=\identity$. The third plaquette then
requires $\Omega X_2^{\dagger} \Omega X_2^{\dagger}=\identity$. Finally the
fourth plaquette yields $\Omega X_4^{\dagger}\Omega X_4^{\dagger} =
\identity$. Therefore we see that this theory space has the same relations
and consequently the same low energy physics as the theory space of 
 Fig. \ref{Fig: moose}. The spectrum of this theory can also be understood
by expanding the plaquettes around the vacuum which we choose to be at $X_i
= \identity$.  Using Eq. \ref{Eq: Omega}, 
we can see the plaquettes give mass to three combinations of triplets and
singlets and to one combination of doublets. One triplet, one singlet and one doublet
scalar are eaten by the $SU(3)$ gauge field multiplet that pick up a mass
and we are left at low energy with two doublet zero modes. 
These are the little Higgs of
our theory and they pick up a negative mass squared through top Yukawa
interaction which can be implemented as in Sec. \ref{Sec: Fermions}. 
There is a large stabilizing quartic interaction which is guaranteed by the 
potential and can be tied to the top quark Yukawa coupling in the
manner described in Sec. \ref{Sec: Yuk2Plaq}.  At the TeV
scale, the theory  contains one doublet, three triplet and three 
singlet scalars 
and one multiplet of $SU(3)$ vector bosons. It also
contains heavy fermions that were introduced in order to cancel the
quadratic divergence associated with the top Yukawa coupling. Because the
top Yukawa is in general larger than the gauge couplings and quartic
interactions, these heavy fermions will
typically be the lightest of the new TeV scale particles.

\section{Conclusion}

The stability of the weak scale requires new physics at the TeV
scale. This physics could be strongly coupled as in technicolor models or
weakly coupled as in supersymmetry. There is now a new class of models that
stabilize the weak scale with weakly coupled new physics qualitatively
different than supersymmetry \cite{Arkani-Hamed:2001nc,Coset}. Higgs bosons in
these theories are pseudo-Goldstone bosons and therefore naturally light. We
studied models of this kind that can be described by general theory
spaces. This generalize the analysis of
\cite{Arkani-Hamed:2001nc,Arkani-Hamed:2002pa} which used toroidal theory
spaces. The physics however remains the same: the Higgs are
pseudo-Goldstone bosons and have their mass protected by approximate chiral
symmetries. The quadratic divergences caused by couplings of the Higgs to
particles of the low energy theory are softened at the TeV scale by ``partners''
of the same spin. The theory remains perturbative up to scales of 
$\sim 10$ TeV where an ultraviolet completion is needed. 

The main result of this paper is the development of systematic procedures
for extracting the low energy particle content and potential form arbitrary
theory spaces and for building theory spaces that produce arbitrary low
energy field content and potential. The former consists in calculating the
classical moduli space of the theory by first gauge fixing and then minimizing
each plaquette, and is equivalent to calculating the fundamental group of
the theory space. We thus learn that the low energy physics of a theory space is
determined by its topology, and different theory spaces with
the same first homotopy group will have the same low energy physics. They
will differ only in their TeV scale spectrum. 

We also derived two simple properties that a theory space must satisfy in
order to be free of quadratic divergences at one loop. 
This put some mild constraints
on the shape of admissible theory spaces. We also showed a simple way of
introducing the top Yukawa coupling without reintroducing quadratic divergences.
The one loop constraints make for minimal TeV scale physics predictions.
To solve the one loop gauge quadratic divergence there must be a $W'$
and $B'$ massive vectors in the \mbox{1 -- 2 TeV} range.   To remove the one
loop quadratic divergences from the tree level scalar potential, 
there must be at least pair of triplets and a pair of singlets
in the \mbox{100 GeV -- 1 TeV} range and an additional set of triplet, doublet
of singlet scalars in the \mbox{1 -- 2 TeV} range.  Finally, for the top quark
coupling, a coloured Dirac fermion in the \mbox{700 GeV -- 1 TeV} range is
necessary.  The lack of striking experimental signatures in the
\mbox{100 -- 500 GeV} range is the surprising feature of this class of models.
In particular, distinguishing this set of models from supersymmetric
models from the two light doublets would be a challenging task at the
Tevatron or LHC.

Finally, we made use of the presence of a site of reduced gauge symmetry
and introduced generalized plaquettes that are gauge invariant but break the
chiral symmetries (the ``$\mathbf{T_8}$ plaquette'' of
\cite{Arkani-Hamed:2001nc,Arkani-Hamed:2002pa}). This allowed us to push to
the TeV scale the
light singlet and triplet scalars that were present before
\cite{Arkani-Hamed:2001nc,Arkani-Hamed:2002pa,2sites} and were the
``$SU(3)$ companions'' of the Higgs doublets. Using these
generalized plaquette we built a minimal model of electroweak symmetry
breaking from theory
space. It is very similar to the model of \cite{2sites} but with the light
triplet and singlet scalar lifted to the TeV scale. In the  100 GeV
region the model has only two Higgs doublets and in the TeV
range has three singlet and triplet scalars, one doublet scalar, one
$SU(3)$ vector boson multiplet and one coloured fermion.

Little Higgs theories are still largely unexplored and there are a lot of model
building and phenomenological studies to be done. Interesting
possibilities include combining the ideas of \cite{Dimopoulos:2002mv} with
little Higgs, pushing the cutoff to higher energies by using a ``cascade''
of theory spaces, detailed studies of collider signatures, and cosmological implications.

\section*{Acknowledgments}
We wish to thank Nima Arkani-Hamed and Andrew Cohen for very valuable
insights . We also thank Martijn Wijnholt and Andrew Neitzke for helpful
discussions. This work is supported in part by the Department of Energy
under Contracts DE-AC03-76SF00098 and the
National Science Foundation under grant PHY-95-14797. T. Gregoire is also
supported by an NSERC fellowship.

\providecommand{\href}[2]{#2}\begingroup\raggedright

\endgroup
\end{document}